\documentclass[fleqn,12pt,twoside]{article}
\usepackage[headings]{espcrc1}
\usepackage{graphics,graphicx,amsmath,array,hhline,fancyhdr}

\begin{document}

\title{Universal scaling of the rapidity dependent elliptic flow
and the perfect fluid at RHIC\footnote{This work was supported by
the OTKA T038406 grant}}
\author{M. Csan\'ad$^{\, 1}$, T. Cs\"org\H{o}$^{\, 2}$, B.
L\"orstad$^{\, 3}$ and A. Ster$^{\, 2}$\\
    {\small $^1$ Department of Atomic Physics, ELTE, Budapest, P\'azm\'any P. 1/A, H-1117} \\
  {\small $^2$ MTA KFKI RMKI, H - 1525 Budapest 114, P.O.Box 49, Hungary}\\
  {\small $^3$ Department of Physics, University of Lund, S-22362 Lund,
  Sweden}}

\maketitle

\runtitle{Universal scaling of the rapidity dependent elliptic
flow and ...}

\runauthor{M. Csan\'ad, T. Cs\"org\H{o}, B. L\"orstad and A. Ster}

\begin{abstract}
The pseudo-rapidity dependence of the elliptic flow at various
excitation energies measured by the PHOBOS Collaboration in Au+Au
collisions at RHIC is one of the surprising results that has not
been explained before in terms of hydrodynamical models. Here we
show that these data are in agreement with theoretical predictions
based on perfect fluid hydrodynamics. We also show that these
PHOBOS data satisfy the universal scaling relation predicted by
the Buda-Lund hydrodynamical model, based on exact solutions of
perfect fluid hydrodynamics.
\end{abstract}

\section{Introduction}

One of the unexpected results from experiments at the Relativistic
Heavy Ion Collider (RHIC) is the relatively strong second harmonic
moment of the transverse momentum distribution, referred to as the
elliptic flow. Measurements of the elliptic flow by the PHENIX,
PHOBOS and STAR collaborations (see
refs.~\cite{Back:2004zg,Adler:2003kt,Adler:2001nb,Sorensen:2003wi})
reveal rich details in terms of its dependence on particle type,
transverse and longitudinal momentum variables, on the centrality
and the bombarding energy of the collision. In the soft transverse
momentum region, these measurements at mid-rapidity are reasonably
well described by hydrodynamical
models~\cite{Adcox:2004mh,Adams:2005dq}. However, the dependence
of the elliptic flow on the longitudinal momentum variable
pseudo-rapidity and its excitation function has resisted
descriptions in terms of hydrodynamical models.

Here we show that these data are consistent with the theoretical
and analytic predictions that are based on perfect fluid
hydrodynamics.

\section{Results}

Our tool in describing the pseudorapidity-dependent elliptic flow
is the Buda-Lund hydrodynamical model. The Buda-Lund hydro
model~\cite{Csorgo:1995bi} is successful
 in describing the BRAHMS, PHENIX, PHOBOS
and STAR data on identified single particle spectra and the
transverse mass dependent Bose-Einstein or HBT radii as well as
the pseudorapidity distribution of charged particles in Au + Au
collisions both at $\sqrt{s_{\rm{NN}}} = 130 $
GeV~\cite{ster-ismd03} and at $\sqrt{s_{\rm{NN}}} = 200 $
GeV~\cite{mate-warsaw03}. However the elliptic flow would be zero
in an axially symmetric case, so we developed the ellipsoidal
generalization of the model that describes an expanding ellipsoid
with principal axes $X$, $Y$ and $Z$. Their derivatives with
respect to proper-time (expansion rates) are denoted by $\dot X$,
$\dot Y$ and  $\dot Z$.

The generalization goes back to the original one, if the
transverse directed principal axes of the ellipsoid are equal, ie
$X=Y$ (and also $\dot X=\dot Y$).

One can define a deviation from axial symmetric flow, a
momentum-space eccentricity:
\begin{equation}
\epsilon_p = \frac{\dot X^2 - \dot Y^2}{\dot X^2 + \dot Y^2}.
\end{equation}

The exact analytic solutions of hydrodynamics (see
ref.~\cite{hidro1,hidro2,hidro3}), which form the basis of the
Buda-Lund hydro model, develop Hubble-flow for late times, ie $X
\rightarrow_{\tau \rightarrow \infty} \dot X \tau$, so the
momentum-space eccentricity $\epsilon_p$ nearly equals space-time
eccentricity $\epsilon$. Hence, in this paper we extract
space-time eccentricity ($\epsilon$) and average transverse flow
($u_t$) from the data, instead of $\dot X$ and $\dot Y$.

In the time dependent hydrodynamical solutions, these values
evolve in time, however, it was show in ref.~\cite{hidro4} that
$\dot X$ and $\dot Y$, and so $\epsilon$ and $u_t$ become
constants of the motion in the late stages of the expansion.

The result for the elliptic flow (under certain conditions
detailed in ref~\cite{bl-ell}) is the following simple universal
scaling law:
\begin{equation}
v_2=\frac{I_1(w)}{I_0(w)}.\label{e:v2w}
\end{equation}

The model predicts an \emph{universal scaling:} every $v_2$
measurement is predicted to fall on the same \emph{universal}
scaling curve $I_1/I_0$ when plotted against $w$.

This means, that $v_2$ depends on any physical parameter
(transverse or longitudinal momentum, center of mass energy,
centrality, type of the colliding nucleus etc.) only trough the
scaling paremeter $w$.

Here $w$ is the scaling variable, defined by
\begin{equation}w=\frac{p_t^2}{4
   \overline{m}_t} \left(\frac{1}{T_{*,y}}
   -\frac{1}{T_{*,x}}\right),
\end{equation}
and
\begin{eqnarray}
   T_{*,x}&=&T_0+\overline{m}_t \, \dot X^2
       \frac{T_0}{T_0 +\overline{m}_t a^2},\\
   T_{*,y}&=&T_0+\overline{m}_t \, \dot Y^2
     \frac{T_0}{T_0 +\overline{m}_t a^2},
\end{eqnarray}
  and
\begin{equation}
    \overline{m}_t = m_t \cosh(\eta_{s}-y).
\end{equation}
Here $a=\langle \Delta T/T \rangle_t$ represents the temperature
gradient in the transverse direction, at the freeze-out, $m_t$ is
the transverse mass, $T_0$ the central temperature at the
freeze-out, while $\eta_{s}$ is the space-time rapidity of the
saddle-point (point of maximal emittivity). This saddlepoint
depends on the rapidity, the longitudinal expansion, the
transverse mass and on the central freeze-out temperature:
\begin{equation}
\eta_{s}-y = \frac{y}{1+\Delta \eta \frac{m_t}{T_0}},
\end{equation}
where $y = 0.5 \log(\frac{E + p_z}{E - p_z})$ is the rapidity and
$\Delta \eta$ represents the elongation of the source expressed in
units of space-time rapidity. See more details in
ref.~\cite{bl-ell}.

Eq.~\ref{e:v2w} depends, for a given centrality class, on rapidity
$y$ and transverse mass $m_t$. Before comparing our result to the
$v_2(\eta)$ data of PHOBOS, we thus performed a saddle point
integration in the transverse momentum variable to end up with a
formula that can be directly fitted to $v_2(\eta)$ of PHOBOS.

The fitting package is available at ref.~\cite{blcvs}, more about
the results (eg. contour plots) are available at
ref.~\cite{blweb}.

\begin{figure}
  \includegraphics[width=450pt]{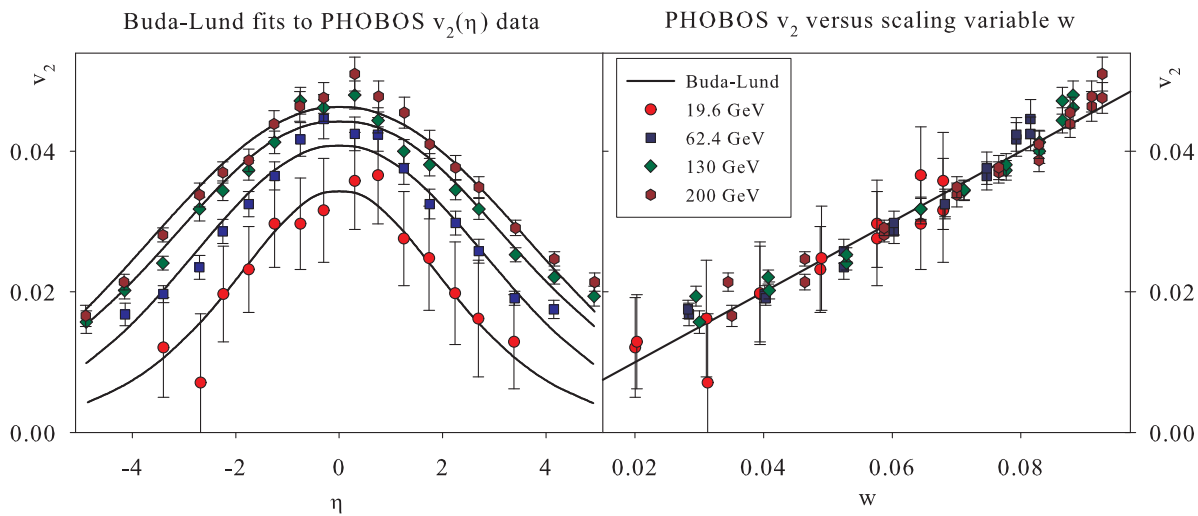}\\
  \caption{On the left: PHOBOS data on the pseudorapidity dependence
  of the elliptic flow~\cite{Back:2004zg}, at various center of mass
  energies, with Buda-Lund fits. On the right: elliptic flow versus
  scaling variable $w$ is plotted. The data points show the predicted
  scaling behavior}
  \label{f:fit}
\end{figure}

We have found that the essential fit parameters are $\epsilon$ and
$\Delta \eta$, and the quality of the fit is insensitive to the
precise value of $T_0$, $a$, $u_t$ and $R_g$. These parameters
dominate the azimuthal-averaged single particle spectra as well as
the HBT (Bose-Einstein) radii, however they only marginally
influence $v_2$. In a broad region their precise value is
irrelevant and does not influence the confidence level of the
$v_2(\eta)$ fits. Hence we have fixed their values as given in the
caption of table~\ref{f:fit}. We also removed points with large
rapidity from the fits in case of lower center of mass energies.

Fits to PHOBOS data of ref.~\cite{Back:2004zg} are shown on the
left panel of fig.~\ref{f:fit}. The right panel of
fig.~\ref{f:fit} demonstrates that these data points follow the
theoretically predicted predicted scaling law.

\begin{table}[ht]
\begin{tabular}{|l|c|c|c|c|}
\hline
                  & 19.6 GeV & 62.4 GeV & 130 GeV & 200 GeV \\
\hline
  $\epsilon$ & 0.294 $\pm$ 0.029  & 0.349 $\pm$ 0.008  & 0.376 $\pm$ 0.005  & 0.394 $\pm$ 0.006 \\
  $\Delta \eta$ & 1.70  $\pm$ 0.25   & 2.16  $\pm$ 0.05   & 2.46  $\pm$ 0.04   & 2.56  $\pm$ 0.04  \\
  $\chi^2$/NDF  & 1.8418/11          & 20.1388/13         & 34.7935/15         & 27.4865/15        \\
  conf. level   & 99.995\%           & 21.4036\%          & 1.00341\%          & 7.03121\%         \\
\hline
\end{tabular}
\begin{center}
  \caption{Results of fits to PHOBOS data of ref.~\cite{Back:2004zg}. Remaining parameters were fixed
  as follows: $T_0=175$ MeV, $a=1.19$ and $u_t=1.64$.}\label{t:fit}
  \end{center}
\end{table}

\section{Conclusions}

In summary, we have shown that the excitation function of the
pseudorapidity dependence of the elliptic flow in Au+Au collisions
is well described with the formulas that are predicted by the
Buda-Lund type of hydrodynamical calculations.

We have provided a quantitative proof of the validity of the
perfect fluid picture of soft particle production in Au+Au
collisions at RHIC but also show here that this perfect fluid
extends far away from mid-rapidity.

The universal scaling of PHOBOS $v_2(\eta)$, expressed by
eq.~\ref{e:v2w}, and illustrated by fig.~\ref{f:fit} provides a
successful quantitative as well as qualitative test for the
appearence of a perfect fluid in Au+Au collisions at various
colliding energies at RHIC.


\begin{thebibliography}{19}
\bibitem{Back:2004zg}
  B.~B.~Back {\it et al.}  [PHOBOS Collaboration],
  Phys.\ Rev.\ Lett.\  {\bf 94}, 122303 (2005)

\bibitem{Adler:2003kt}
  S.~S.~Adler {\it et al.}  [PHENIX Collaboration],
  Phys.\ Rev.\ Lett.\  {\bf 91}, 182301 (2003)

\bibitem{Adler:2001nb}
  C.~Adler {\it et al.}  [STAR Collaboration],
  Phys.\ Rev.\ Lett.\  {\bf 87}, 182301 (2001)

\bibitem{Sorensen:2003wi}
  P.~Sorensen  [STAR Collaboration],
  J.\ Phys.\ G {\bf 30}, S217 (2004)

\bibitem{Adcox:2004mh}
  K.~Adcox {\it et al.}  [PHENIX Collaboration],
  Nucl.\ Phys.\ A {\bf 757}, 184 (2005)

\bibitem{Adams:2005dq}
  J.~Adams {\it et al.}  [STAR Collaboration],
  Nucl.\ Phys.\ A {\bf 757}, 102 (2005)

\bibitem{Csorgo:1995bi}
T.~Cs\"org\H{o} and B.~L\"orstad,
Phys.\ Rev.\ C {\bf 54} (1996) 1390

\bibitem{ster-ismd03}
M. Csan\'ad, T. Cs\"org\H{o}, B. L\"orstad, A. Ster,
Act. Phys. Pol. B{\bf 35}, 191 (2004)

\bibitem{mate-warsaw03}
M.~Csan\'ad, T.~Cs\"org\H{o}, B.~L\"orstad and A.~Ster,
Nukleonika {\bf 49}, S45 (2004)

\bibitem{hidro1}
  S.~V.~Akkelin, T.~Cs\"org\H{o}, B.~Luk\'acs, Y.~M.~Sinyukov, M.~Weiner,
  PLB {\bf 505}, 64 (2001)

\bibitem{hidro2}
  T.~Cs\"org\H{o},
  hep-ph/0111139.

\bibitem{hidro3}
  T.~Cs\"org\H{o}, L.~P.~Csernai, Y.~Hama and T.~Kodama,
  Heavy Ion Phys.\ A {\bf 21}, 73 (2004)

\bibitem{hidro4}
  T.~Cs\"org\H{o} and J.~Zim\'anyi,
  Heavy Ion Phys.\  {\bf 17}, 281 (2003)

\bibitem{bl-ell}
M.~Csan\'ad, T.~Cs\"org\H{o} and B.~L\"orstad,
  Nucl.\ Phys.\ A {\bf 742}, 80 (2004)

\bibitem{blcvs}
\verb|http://www.phenix.bnl.gov/viewcvs/offline/analysis/budalund/|

\bibitem{blweb}
\verb|http://csanad.web.elte.hu/phys/v2eta|

\end{thebibliography}
\end{document}